\documentclass[12pt]{article}
\usepackage{cite}
\usepackage{graphicx}
\usepackage{amssymb,amsfonts}
\usepackage{hangcaption}

\renewcommand{\theequation}
{\arabic{section}.\arabic{equation}}

\makeatletter
\def\eqnarray{ \stepcounter{equation} \let\@currentlabel=\theequation
 \global\@eqnswtrue
 \global\@eqcnt\z@
 \tabskip\@centering
 \let\\=\@eqncr
 $$\halign to \displaywidth\bgroup\@eqnsel\hskip\@centering
 $\displaystyle\tabskip\z@{##}$&\global\@eqcnt\@ne
 \hfil$\displaystyle{{}##{}}$\hfil
 &\global\@eqcnt\tw@$\displaystyle\tabskip\z@{##}$\hfil
 \tabskip\@centering&\llap{##}\tabskip\z@\cr}
\makeatother

\makeatletter
\def\@arrayacol{\edef\@preamble{\@preamble \hskip .5\arraycolsep}}
\def\array{\let\@acol\@arrayacol \let\@classz\@arrayclassz
\let\@classiv\@arrayclassiv \let\\\@arraycr\def\@halignto{}\@tabarray}
\makeatother

\makeatletter
\newcounter{subeqncnt}
\def\thesubeqncnt{\alph{subeqncnt}}
\def\subequations{\begingroup%
   \stepcounter{equation}\edef\@tempa{\theequation}%
   \let\c@equation\c@subeqncnt\c@subeqncnt\z@
   \edef\theequation{\@tempa\noexpand\thesubeqncnt}}

\makeatother

\newcommand{\captionfonts}{\small}
\makeatletter 
\long\def\@makecaption#1#2{%
\vskip\abovecaptionskip
\sbox\@tempboxa{{\captionfonts #1: #2}}%
\ifdim \wd\@tempboxa >\hsize {\captionfonts #1: #2\par} \else
\hbox to\hsize{\hfil\box\@tempboxa\hfil}%
\fi \vskip\belowcaptionskip}
\makeatother 

\newcommand{\del}{\partial}

\newcommand{\dd}{{\rm d}}

\def\imo{i}

\begin{document}

\setlength{\baselineskip}{7mm}

\begin{flushright}
{\tt SHU-Pre2009-01} \\
\end{flushright}

\vspace{1cm}

\begin{center}
{\Large Shear viscosity, instability and  the upper bound of the
Gauss-Bonnet coupling constant}

\vspace{1cm}

{\sc{Xian-Hui Ge}}$^*$, {\sc{Sang-Jin Sin}}$^{\dagger }$

$*${\it{Department of Physics, Shanghai University},} \\
{\it{Shanghai 200444, China}} \\
{\sf{gexh@shu.edu.cn}}
\\
$\dagger$ {\it{Department of Physics,}} {\it{Hanyang University,}}
{\it{Seoul 133-791, Korea}} \\
{\sf{sjsin@hanyang.ac.kr}}
\end{center}

\vspace{1.5cm}

\begin{abstract}
 We compute the dimensionality dependence of $\eta/s$ for charged
black branes with Gauss-Bonnet correction. We find that  both
causality and stability constrain the value of Gauss-Bonnet coupling
constant to be bounded by $1/4$ in the infinite dimensionality
limit. We further show that higher dimensionality stabilize the
gravitational perturbation. The stabilization of the perturbation in
higher dimensional space-time is a straightforward consequence of
the Gauss-Bonnet coupling constant bound.
\end{abstract}

\section{Introduction}
\setcounter{equation}{0} \setcounter{footnote}{0} The AdS/CFT
correspondence \cite{ads/cft,gkp,w} provides an interesting
theoretical framework for studying relativistic hydrodynamics of
strongly coupled gauge theories. The result of RHIC experiment on
the viscosity/entropy ratio turns out to be in favor of the
prediction of AdS/CFT ~\cite{pss0,kss,bl}. Some attempt has been
made to map the entire process of RHIC experiment in terms of
gravity dual \cite{ssz}. The way to include chemical potential in
the theory was figured out in~\cite{ksz,ht}. Phases of these
theories were also discussed in \cite{nssy1,kmmmt,nssy2,huang,ht}.

It had been conjectured that the viscosity value of theories with
gravity dual may give a lower bound for the $\eta/s=\frac{1}{4 \pi}$
for all possible liquid \cite{kovtun}. However, in the presence of
higher-derivative gravity corrections, the viscosity bound and
causality are also violated as a consequence
\cite{kp,shenker,shenker1,cai1,neupane,gmsst,cai2,cai3}.

The higher derivative gravity terms are related to the (in)stability
issues of black holes. The black hole stability issues are a crucial
problem because black hole solutions are no longer unique in
spacetime with higher than four dimensions. The instability of
$D$-dimensional asymptotically flat Einstein-Gauss-Bonnet black
holes has been discussed by several authors\cite{dotti,konoplya}.
Their results show that for the gravitational perturbations of
Schwarzschild black holes in $D\geq 5 $ Gauss-Bonnet gravity, the
instability occurs only for $D=5$ and $D=6$ cases at large value of
$\alpha'$ \cite{konoplya}. In the previous paper \cite{gmsst}, we
computed the charge dependence of $\eta/s$ for Gauss-Bonnet theory
and noticed that charges introduced instability of the black brane
even in the range $0<\lambda\leq 0.09$.

The purpose of this paper is to perform a complete computation of
$\eta/s$ including the charge and Gauss-Bonnet correction to an
arbitrary dimensionality , and to determine the causality and
stability constraints on the parameters of the black hole. Both the
causality and stability constraints give the same result that
$\lambda$ should be bounded by $1/4$ for an arbitrary high
dimensionality.
 We further find that higher
dimensionality stabilize the tensor type perturbation.

\section{Viscosity to entropy density ratio}
We have explored the charge dependence of $\eta/s$ in the presence
of Gauss-Bonnet term for five-dimensional AdS black branes
\cite{gmsst}. In this section, we generalize the previous result on
$\eta/s$ \cite{gmsst} to $D$-dimensional cases. Let us start by
introducing the following action in $D$ dimensions which includes
Gauss-Bonnet terms and $U(1)$ gauge field:
\begin{equation}
\label{action} I=\frac{1}{16 \pi G_{D}}\!\int\!\dd^{D}\!x
\sqrt{-g}\Big(R-2\Lambda+\alpha'\left(R_{\mu\nu\rho\sigma}
R^{\mu\nu\rho\sigma}-4R_{\mu\nu}R^{\mu\nu}+R^2\right)-4 \pi G_{D}
F_{\mu\nu}F^{\mu\nu}\Big),
\end{equation}
where $\alpha'$ is a (positive) Gauss-Bonnet coupling constant with
dimension $\rm(length)^2$ and the field strength is defied as
$F_{\mu\nu}(x)=\del_\mu A_\nu(x)-\del_\nu A_\mu (x)$. The
thermodynamics and geometric properties of black objects in
Gauss-Bonnet gravity were studied in several papers
\cite{g1,g2,g3,cvetic,g4,gs,ast}.

The charged black brane solution in $D$ dimensions for this action
is described by~\cite{cvetic}
\begin{subequations}
\begin{eqnarray}
\label{metric} \dd s^2 &=& \displaystyle -H(r)N^2\dd
t^2+H^{-1}(r)\dd r^2+\frac{r^2}{l^2} \dd x^{i}\dd x^{j},
\\
A_t &=& \displaystyle -\frac{Q}{4\pi(D-3)r^{D-3}},
\end{eqnarray}
\end{subequations}
with
\begin{eqnarray}
H(r)&=& \frac{r^2}{2
\alpha}\left[1-\sqrt{1-\frac{4\alpha}{l^2}\bigg(1-\frac{ml^2}{r^{D-1}}+\frac{q^2l^2}{r^{2D-4}}}\bigg)
\right],\nonumber\\&=&\frac{r^2}{2\lambda
l^2}\left[1-\sqrt{1-4\lambda\bigg(1-\frac{r^{D-1}_{+}}{r^{D-1}}
-a\frac{r^{D-1}_{+}}{r^{D-1}}+a\frac{r^{2D-4}_{+}}{r^{2D-4}}\bigg)}\right],
\nonumber\\
\Lambda &=&-\frac{(D-1)(D-2)}{2l^2}.
\end{eqnarray}
where  $\alpha$ and $\alpha'$ are connected by a relation
$\alpha=(D-4)(D-3)\alpha'$, $\lambda=\alpha/l^2$,
$a=\frac{q^2l^2}{r^{2D-4}_{+}}$ and the parameter $l$ corresponds to
AdS radius. The horizon is located at $r=r_{+}$. The gravitational
mass $M$ and the charge $Q$ are expressed as
\begin{eqnarray*}
M&=&\frac{(D-2)V_{D-2}}{16 \pi G_D  }m,
\\
Q^2&=&\frac{2\pi (D-2)(D-3)}{ G_D  }q^2.
\end{eqnarray*}

Taking the limit $\alpha'\rightarrow 0$, the solution corresponds to
one for Reissner-Nordstr\"om-AdS (RN-AdS). The hydrodynamic analysis
in this background has been done in\cite{gmsst1,msst}.

The constant $N^2$ in the metric (\ref{metric}) can be fixed at the
boundary whose geometry would reduce to flat Minkowski metric
conformaly, i.e.\ $\dd s^2\propto -c^2\dd t^2+\dd\vec{x}^2$. On the
boundary $r\rightarrow\infty$, we have
$$
H(r)N^2 \rightarrow\frac{r^2}{l^2},
$$
so that $N^2$ is found to be
\begin{equation}
N^2=\frac{1}{2}\Big(1+\sqrt{1-4 \lambda}\ \Big).\label{N}
\end{equation}
Note that the boundary speed of light is specified to be unity
$c=1$. Eq.(\ref{N}) implies that the significant value of $\lambda$
lies in the region $\lambda\leq 1/4$. We will confirm this result
from the causality and stability analysis in section 3 and 4.

The temperature at the event horizon is defined as
\begin{equation}
T=\frac{1}{2\pi\sqrt{g_{rr}}}\frac{\dd \sqrt{g_{tt}}}{\dd
r}=\frac{Nr_+}{4\pi l^2}\left((D-1)-(D-3)a\right).
\end{equation}
 The black brane approaches
extremal when $a\rightarrow \frac{D-1}{D-3}$ (i.e.\ $T\rightarrow
0$). The entropy density is given by \cite{g3}
\begin{equation}
s=\frac{1}{4 G_{D}}\frac{r^{D-2}_{+}}{l^{D-2}}.
\end{equation}
We will calculate the shear viscosity of the boundary theory using
the Kubo formula
\begin{equation}
\eta=\lim_{\omega\rightarrow 0}\frac{1}{2 \omega}\int dt
d\vec{x}\mbox{e}^{-i\omega
t}<[T_{xy}(x),T_{xy}(0)]>=-\lim_{\omega\rightarrow
0}\frac{1}{\omega} {\rm Im}G_{xy,xy}(\omega,0), \label{kubo}
\end{equation}
where $G(\omega,0)$ is the  retarded Green¡¯s function for $T_{xy}$:
\begin{equation}
\label{green}
G_{xy,xy}(\omega,k)=-i\int d t d{ x} \mbox{e}^{ik\cdot
x+i\omega t}\theta(t)<[T_{xy}(x),T_{xy}(0)]>
\end{equation}
It is convenient to introduce coordinate in the following
computation
\begin{eqnarray}
&&z=\frac{r}{r_{+}},
~~\omega=\frac{l^2}{r_{+}}\bar{\omega},~~k_{3}=\frac{l^2}{r^2_{+}}\bar{k}_3,~~
f(z)=\frac{l^2}{r^2_{+}}H(r), \nonumber\\
&&f(z)=\frac{z^2}{2
\lambda}\bigg[1-\sqrt{1-4\lambda\bigg(1-\frac{a+1}{z^{D-1}}+\frac{a}{z^{2D-4}}}\bigg)\bigg]
\end{eqnarray}
We now study the tensor type perturbation
$h^{x}_{y}(t,x_3,z)=\phi(t,x_3,z)$ on the black brane background of
the form
$$
ds^2=-f(z)N^2\dd t^2+\frac{\dd z^2}{b^2 f(z)}+\frac{z^2}{b^2
l^2}\left(2\phi(t,x_3,z)\dd x \dd y+\sum^{D-2}_{i=1}\dd
x^2_{i}\right),
$$
where $b=\frac{1}{r^2_{+}}$.
 Using Fourier decomposition
$$
\phi(t, x_3,z) = \!\int\!\frac{\dd^{D-1}k}{(2\pi)^{D-1}}
\mbox{e}^{-i\bar{\omega} t+i\bar{k}_{3}x_3}\phi(k, z),
$$
we can obtain the following linearized equation of motion for
$\phi(z)$ from the Einstein-Gauss-Bonnet-Maxwell field equation:
\begin{equation}\label{maineq}
g(z) \phi''+g'(z)\phi'+g_2 \phi=0
\end{equation}
where
\begin{eqnarray}
&&g(z)=z^{D-2}f\left\{1-\frac{2 \lambda}{D-3}
\left[z^{-1}f'+z^{-2}(D-5)f\right]\right\}\nonumber
\\
&&g_2=g(z)\frac{\omega^2}{N^2f^2}-k^2_{3}z^{D-4}\times\nonumber
\\
&&\left[1-\frac{2\lambda}{(D-3)(D-4)}\left(f''+(D-5)(D-6)z^{-2}f+2(D-5)z^{-1}f'\right)\right],
\end{eqnarray}
and the prime denotes the derivative with respect to $z$.

For the convenient calculation of the shear viscosity, we further
introduce a new variable $u(=\frac{1}{z})$. Then we solve the
equation of motion for transverse graviton Eq.(\ref{maineq}) in
hydrodynamic regime i.e.\ small $\omega$ and $k$. The solution to
Eq.(\ref{maineq}) as
\begin{equation}
\phi(z)=(1-u)^\nu F(u), \label{solF}
\end{equation}
where $F(u)$ is a regular function at the horizon $u=1$, so that the
singularity at the horizon might be extracted. The parameter $\nu$
can be fixed as $\nu=\pm i\omega/4\pi T$ by substituting
Eq.(\ref{solF}) into the equation of motion. Usually we choose
\begin{equation}
\nu=-i\frac{\omega}{4\pi T}
\end{equation}
as the incoming wave condition. To obtain the shear viscosity via
Kubo formula (\ref{kubo}), we only need know the $\omega\rightarrow
0$ behavior of the transverse graviton, so it is sufficient to
expand $F(u)$ in terms of frequencies up to the linear order of
$\omega(=i4\pi T\nu)$,
\begin{equation}
F(u) =F_0(u)+\nu F_1(u) + {\cal O}(\nu^2, k^2). \label{series}
\end{equation}
Expanding (\ref{maineq}) to the first order of $\nu$, we get the
following form,
\begin{equation}
\label{eq} \left[g(u)F'(u)\right]'
-\nu\left(\frac{1}{1-u}g(u)\right)'F(u)-\frac{2\nu}{1-u}g(u)F'(u)=0.
\end{equation}
Substituting the series expansion (\ref{series}) into the equation
(\ref{eq}), one can get the equations of motion for $F_0(u)$ and
$F_1(u)$ recursively. Following the procedure given in \cite{gmsst},
we easily get
\begin{equation}
F_0(u)=C, \qquad (\mbox{const.}).
\end{equation}
and
\begin{equation}
F'_1 (u)=\frac{C}{1-u}+\frac{C_2}{g(u)}.
\end{equation}
The integration constant $C_2$ can be fixed by the regularity
condition of $F_1(u)$ at the horizon. So, the regularity condition
at $u=1$ implies
\begin{equation}
C_2=-\bigg[\left((D-1)-(D-3)a\right)(1-\frac{2\lambda}{D-3}
((D-1)-(D-3)a))\bigg]C.
\end{equation}
The remaining constant $C$ is estimated in terms of boundary value
of the field,
$$
\lim_{u\rightarrow 0}\phi(z)=\phi^{(0)},
$$
so that we could fix
\begin{equation}
C=\phi^{(0)}\Big(1+{\cal O}(\nu)\Big).
\end{equation}

Now let us calculate the retarded Green function. Using the equation
of motion, the action reduces to the surface terms. The relevant
part is given as
\begin{equation}
I[\phi(u)]= -\frac{r^{D-1}_{+}N}{16\pi G_D l^D}
\!\int\!\frac{\dd^{D-1} k}{(2\pi)^{D-1}}
\Big(g(u)\phi(u)\phi'(u)+\cdots\Big)\Bigg|_{u=0}^{u=1}.
\end{equation}
Near the boundary $u=\varepsilon$, using the obtained perturbative
solution for $\phi(u)$, we can get
\begin{eqnarray}
\phi'(\varepsilon) &=&
-\nu\frac{\bigg[\left((D-1)-(D-3)a\right)(1-\frac{2\lambda}{D-3}
((D-1)-(D-3)a))\bigg]}{g({\varepsilon})}\phi^{(0)} +{\cal O}(\nu^2,
k^2) \nonumber
\\
&=& i\omega\bigg(\frac{l^2}{4Nr_+}\bigg)
\frac{1-\frac{2\lambda}{D-3}
((D-1)-(D-3)a)}{g(\varepsilon)}\phi^{(0)} +{\cal O}(\omega^2, k^2).
\end{eqnarray}
Therefore we can read off the correlation function from the relation
(\ref{green}),
\begin{equation}
G_{xy \ xy}(\omega, k) =-i\omega\frac{1}{16\pi
G_D}\left(\frac{r^{D-2}_{+}}{l^{D-2}}\right)
\Big(1-\frac{2\lambda}{D-3} [(D-1)-(D-3)a]\Big) +{\cal O}(\omega^2,
k^2),
\end{equation}
where we subtracted contact terms. Then finally, we can obtain the
shear viscosity by using Kubo formula (\ref{kubo}),
\begin{equation}
\eta=\frac{1}{16 \pi
G_D}\left(\frac{r^{D-2}_{+}}{l^{D-2}}\right)\Big(1-\frac{2\lambda}{D-3}
[(D-1)-(D-3)a]\Big).
\end{equation}
The ratio of the shear viscosity to the entropy density is found to
be
\begin{equation}
\frac{\eta}{s} =\frac{1}{4 \pi } \left(1-\frac{2\lambda}{D-3}
[(D-1)-(D-3)a]\right).
\end{equation}
The above result agrees with \cite{gmsst} when $D=5$. Since
$\lambda$ is bounded by $1/4$, the shear viscosity never approaches
zero in higher than $5D$ Gauss-Bonnet theory. When $a=0$ (no
charges), $\eta/s=(1-\frac{2\lambda(D-1)}{(D-3)})/(4\pi)$, we
recover the result in Ref.\cite{shenker}. It is also worth noting
that for extremal case ($a=\frac{D-1}{D-3}$), the ratio of the shear
viscosity to entropy density receives no corrections from
Gauss-Bonnet terms. In the next two sections, we will show
explicitly causality and stability impose more rigorous constraint
on the value of $\lambda$ and the upper bound  of $\lambda$ is
$1/4$.

\section{Causality constraints}
\setcounter{equation}{0} \setcounter{footnote}{0} The authors in
~\cite{shenker, shenker1} demonstrated that  the causality could be
violated if one introduced Gauss-Bonnet terms. In \cite{gmsst}, it
is found that the presence of charge does not contribute to
causality and the result of ~\cite{shenker, shenker1} is universal
for charged black branes. In this section, we investigate the
dimensionality dependence of the causality constraints.

Due to higher derivative terms in the gravity action, the equation
(\ref{maineq}) for the propagation of a transverse graviton differs
from that of a minimally coupled massless scalar field propagating
in the same background geometry. Writing the wave function as
\begin{equation}
\label{phi} \phi(x_3,z)=\mbox{e}^{-i\omega t+ikz+ik_{3}x_{3}},
\end{equation}
and taking large momenta limit $k^\mu\rightarrow\infty$, one can
find that the equation of motion (\ref{maineq}) reduces to
\begin{equation}
\label{effeq} k^{\mu}k^{\nu}g_{\mu\nu}^{\rm eff}\simeq 0,
\end{equation}
where the effective metric is given by
\begin{equation}
\dd s^2_{\rm eff} =g^{\rm eff}_{\mu\nu}\dd x^\mu\dd x^\nu ={N^2
f(r)} \left(-\dd t^2+\frac{1}{c^2_g}\dd x^2_3\right)
+\frac{1}{f(r)}\dd r^2.
\end{equation}
Note that $c^2_g$ can be interpreted as the local speed of graviton:
\begin{equation}
c^2_g(z)=\frac{N^2
f}{z^2}\frac{1-\frac{2\lambda}{(D-3)(D-4)}\left(f''+(D-5)(D-6)z^{-2}f+2(D-5)z^{-1}f'\right)}{1-\frac{2
\lambda}{D-3} \left[z^{-1}f'+z^{-2}(D-5)f^2\right]}.
\end{equation}
We can expand $c^2_g$ near the boundary $z \rightarrow\infty$,
\begin{eqnarray}
c^2_g-1=&&
\left(-\frac{(D^2-5D+10)(1+a)}{2(D-3)(D-4)}\right.\nonumber\\&&
\left.+\frac{(D-1)(1+a)}{(D-3)(D-4)(1-4\lambda)}
-\frac{1+a}{2\sqrt{1-4\lambda}}\right)\frac{1}{z^{D-1}}
+\mathcal{O}(z^{-D}).
\end{eqnarray}
As the local speed of graviton should be smaller than $1$ (the local
speed of light of the boundary CFT), we require
\begin{equation}
-\frac{(D^2-5D+10)}{2(D-3)(D-4)}+\frac{(D-1)}{(D-3)(D-4)(1-4\lambda)}
-\frac{1}{2\sqrt{1-4\lambda}}\leq 0.
\end{equation}
The above formula leads to
\begin{equation}
\lambda_{\rm causality} \leq
\frac{D^4-10D^3+41D^2-92D+96}{4(D^2-5D+10)^2}.
\end{equation}
without any charge dependence. As $D$ is large enough,  the above
formula becomes
\begin{equation}
\lim_{D\rightarrow \infty}\lambda_{\rm causality} \leq \frac{1}{4}.
\end{equation}
We can rewrite the above equation from the relation
$\lambda=(D-3)(D-4)\alpha'/l^2$,
\begin{equation}
\frac{\alpha'}{l^2} \leq
\frac{D^4-10D^3+41D^2-92D+96}{4(D^2-5D+10)^2 (D-3)(D-4)}.
\end{equation}
\noindent
\begin{figure}[htbp]
\begin{center}
\includegraphics*[scale=0.5]{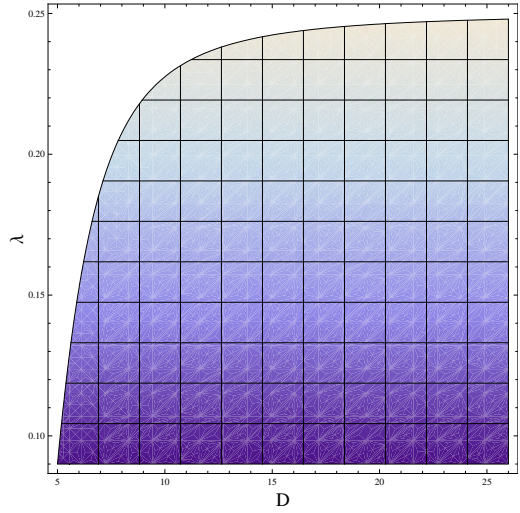}
\end{center}
\caption{The minimal value of $\lambda$ constrained by causality.
The upper bound of $\lambda$ is $1/4$.} \label{lambdabound}
\end{figure}
Figure 1 demonstrates that causality constrains the value of
$\lambda$. When $D=5$, it goes as $\lambda\leq 0.09$ same as the
result of \cite{shenker1} and when $D\rightarrow \infty$, $\lambda$
is bounded by $1/4$.

Now, we rewrite the wave function in a Schr$\ddot{o}$dinger form,
\begin{equation}
-\frac{d^2 \psi}{dr^2_{*}}+V\left(z(r_{*})\right)\psi=\omega^2 \psi,
~~~\frac{dr_{*}}{dz}=\frac{1}{Nf(z)},\label{schr}
\end{equation}
where $\psi\left(z(r_{*})\right)$ and the potential is defined by
\begin{eqnarray}
&&\psi =K(z)\phi,~~~K(z)\equiv\sqrt{\frac{g(z)}{z^{D-2}f(z)}},
V=k^2c^2_g+V_{1}(z),\nonumber\\ &&V_{1}(z)\equiv
N^2\left[\left(f(z)\frac{\partial \ln K(z)}{\partial
z}\right)^2+f(z)\frac{\partial}{\partial z}\left(f(z)\frac{\partial
\ln K(z)}{\partial z}\right)\right]
\end{eqnarray}
From the geodesic  equation of motion
\begin{equation}
g^{\rm eff}_{\mu\nu}\frac{\dd x^{\mu}}{\dd s}\frac{\dd x^{\nu}}{\dd
s}=0,
\end{equation}
and the Bohr-Sommerfield quantization condition
\begin{equation}
 \int \dd r_{\ast}\sqrt{\omega^2-k^2 c^2_g}=(n-\frac{1}{4})\pi,
\end{equation}
one can find that the group velocity of the test particle along the
geodesic line is given by \cite{shenker1}
\begin{equation}
 v_g=\frac{\dd \omega}{\dd k}\rightarrow c_g>1.
\end{equation}
Therefore, signals in the boundary CFT propagate outside of the
light cone and microcausality violation happens (for a more detailed
and explicit discussion on causality violation, see
\cite{shenker1}). Now we can conclude that as dimensions of
space-time go up, causality restricts the value of $\lambda$ in the
region $\lambda\leq 1/4$.  In next section, we will prove that in
the extremal limit $a\rightarrow \frac{D-1}{D-3}$, the stability of
the black brane also requires that  $\lambda$ should also be bounded
by $1/4$.
\section{Stability constraints}
\setcounter{equation}{0} \setcounter{footnote}{0} In \cite{gmsst},
it was found that apart from the causality violation, for RN-AdS
black brane in Gauss-Bonnet theory, the charges give new instability
of the black brane within the window of $0<\lambda\leq 0.09$. Now,
we will show that higher $D$ stabilize the gravitational
perturbation.

\begin{figure}[htbp]
\begin{center}
\includegraphics*[scale=0.5]{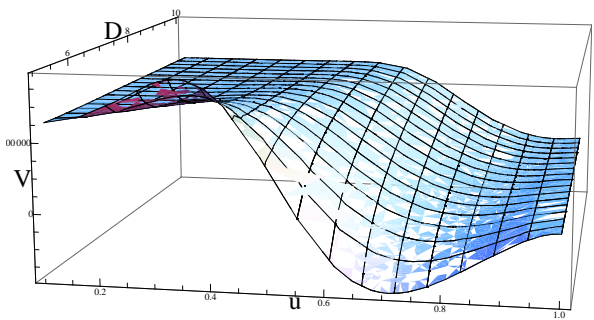}
\end{center}
\caption{Schrodinger potential V(u) as a function of $u$
($u=\frac{1}{z}$) and $D$ for $\lambda=0.24$ and
$a=\frac{D-1}{D-3}$.} \label{potential3D}
\end{figure}

From Figure \ref{potential3D}, we can see that the Schr\"odinger
potential develops a negative gap near the horizon, but the gap
becomes flatter as $D$ increases.
\noindent
We will now show that while in the large momentum limit, the
negative-valued potential leads to instability of the black brane,
higher $D$ tends to suppress those unstable perturbations. In the
large momenta limit $k^\mu\rightarrow \infty$, the dominant
contribution to the potential is given by $k^2 c^2_g$. In
\cite{gmsst}, it was found that for near extremal cases, $c^2_g$ can
be negative near the horizon and $V\simeq k^2 c^2_g$ can be deep
enough (see Figure \ref{potential3D}). Thus bound states can live in
the negative-valued well. The negative energy bound state
corresponds to modes of tachyonic mass on Minkowski slices
\cite{troost} and signals an instability of the black
brane~\cite{dotti,konoplya}. As $D$ goes up, we will see new physics
in the following.   Let us expand $c^2_g$ in series of
$(1-\frac{1}{z})$,
\begin{eqnarray}
c^2_g=&&
\left[(D-3)a-D-1\right]\left(1+\sqrt{1-4\lambda}\right)\left\{D^2
\left[4\lambda^2(a-1)^2+2(a+1)\lambda-1\right]\right.\nonumber\\
&&\left.-D\left[8\lambda^2(3a^2-4a+1)+2\lambda(a+7)\right]+\left[\lambda^2(3a-1)^2-3\lambda(a-1)-3\right]
\right\}\nonumber\\
&&\left\{2 (D-4) \left[-3+(2-6 a) \lambda +D (1+2 (a-1) \lambda
)\right]\right\}^{-1}(1-\frac{1}{z})+\mathcal{O}((1-\frac{1}{z})^2).
\end{eqnarray}
Since $0\leq a\leq \frac{D-3}{D-1}$, and $0\leq \frac{1}{z} \leq 1$,
$c^2_g$ will be negative, if
\begin{eqnarray}
&&\left\{D^2
\left[4\lambda^2(a-1)^2+2(a+1)\lambda-1\right]\right.\nonumber\\
&&\left.-D\left[8\lambda^2(3a^2-4a+1)+2\lambda(a+7)\right]+\left[\lambda^2(3a-1)^2-3\lambda(a-1)-3\right]
\right\}\nonumber\\
&&\left\{2 (-4+D) \left[-3+(2-6 a) \lambda +D (1+2 (-1+a) \lambda
)\right]\right\}^{-1} <0.
\end{eqnarray}
From the above formula, we find the critical value of $\lambda$,
\begin{eqnarray}
\label{4.3}
\lambda_{\rm c}(D,a)&& =\frac{1}{4}\bigg\{-(D-1)(D-6)+(D-3)(D+2)a \nonumber\\
&&+\bigg\{(D-1)^2(5D^2-40D+84)+(D-3)^2(5D^2-24D+52)a^2\nonumber\\
&&-6a(D^2-8D+20)(D-3)(D-1)\bigg\}^{\frac{1}{2}}\bigg\}\bigg\{1+(D-3)a-D\bigg\}^{-2}
.
\end{eqnarray}
\begin{figure}[htbp]
\begin{center}
\includegraphics*[scale=0.5]{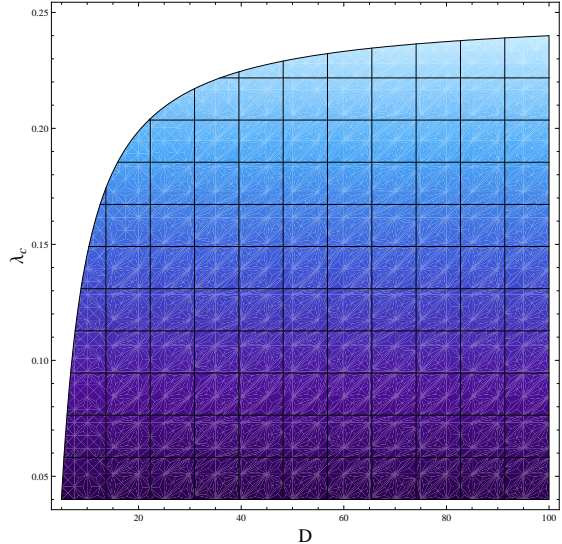}
\end{center}
\caption{The minimal value of $\lambda$ constrained by instability
in the limit  $a\rightarrow \frac{D-1}{D-3}$. The figure shows that
$\lambda_c$ is bounded by $0.25$} \label{lambdac}
\end{figure}
Above the line of $\lambda_{\rm c}$, $c^2_g$ can be negative (see
figure \ref{lambdac}). The minimal value of $\lambda_{\rm c}$ can be
obtained in the limit $a\rightarrow (\frac{D-1}{D-3})$,
\begin{equation}
\label{4.4} \lambda_{\rm c, \ min} =
\frac{1}{4}\frac{(D-3)(D-4)}{(D-1)(D-2)}.
\end{equation}
That is to say
\begin{equation}
\left(\frac{\alpha'}{l^2}\right)_{\rm c, \ min} =
\frac{1}{4(D-1)(D-2)}.
\end{equation}
When $D=5$, $\lambda_{\rm c, \ min}=\frac{1}{24}$, we recover the
result obtained in ref.\cite{gmsst}. Eq(\ref{4.4}) indicates that
for any value of $a$, the quasinormal modes (QNMs) become stable
under the line $\lambda_{\rm c, \ min}$.

 As the value of
$D$ increases, one finds that $\lambda_{\rm c, \ min}$ is also
bounded by $1/4$, i.e.
\begin{equation}
\lim_{(D,a)\rightarrow(\infty,\frac{D-1}{D-3})}\lambda_{\rm
c}=\frac{1}{4}
\end{equation}
Note that this value is obtained in the extremal limit.
 Different from causality violation, the stability of the
black brane depends on the charge. It would be very interesting to
see for fixed value of charge, for which value of $\lambda$ the QNMs
become stable. Eq(\ref{4.3}) tells us that for $\lambda<\lambda_{\rm
c}(D,a)$, the black brane is always stable. Actually, as pointed out
in \cite{gmsst}, for fixed $D$ the two lines $\lambda_{\rm c}(a)$
and $\lambda_{\rm causality}$ separates the physics into four
regions in $(a,\lambda)$ space: consistent region; only causality
violation region; only unstable modes region; causality violation
and unstable modes region(see figure 4 in \cite{gmsst} for more
details).
 \begin{table*}[htbp]
\begin{center}
\begin{tabular}{|c|c|c|c|c|c|c|c|c|}
\hline
$D$&$\lambda=0.20$&$\lambda=0.18$&$\lambda=0.16$&$\lambda=0.12$&$\lambda=0.10$\\
\hline
$6$&$ 122.7276\imo$&$ 92.5503\imo$&$  66.0012\imo$&$18.5027\imo$&$ \rm - $\\
$7$&$ 75.3435 \imo$&$52.8819\imo$&$ 32.5866\imo$&$\rm - $&$\rm -$\\
$8$&$ 49.7368 \imo$&$ 30.9208\imo$&$16.8584$&$ \rm -$&$ \rm -$\\
$9$&$ 34.7918\imo$&$ 20.0792\imo$&$7.5717 \imo$&$ \rm -$&$ -$\\
$10$&$ 24.7602\imo$&$ 11.9079\imo$&$\rm -$&$ \rm -$&$ -$\\
\hline
\end{tabular}
\caption{Unstable QNMs for charged GB black brane perturbation of
tensor type for fixed charge ($a=1.20$) and $k_3=500$.}\label{one}
\end{center}
\end{table*}
\begin{table*}[htbp]
\begin{center}
\begin{tabular}{|c|c|c|c|c|c|c|c|c|}
\hline
$D$&$a=1.4$&$a=1.2$&$a=1.0$&$a=0.8$&$a=0.6$&$a=0.4$\\
\hline
$6$&$ 129.001\imo$&$122.7276 \imo$&$  110.9159\imo$&$91.1455\imo$&$59.3517\imo$&$7.0564\imo$\\
$7$&$ 79.3218 \imo$&$75.3435\imo$&$ 66.5505\imo$&$ 37.3177\imo$&$ 24.3269\imo$&$ \rm - $\\
$8$&$  70.0214\imo$&$ 49.7368\imo$&$ 42.0717\imo$&$ 26.6617\imo$&$ -$&$ \rm -$\\
$9$&$ \rm -$&$ 32.6603\imo$&$27.1475\imo$&$ 12.6057\imo$&$ -$&$ -$\\
$10$&$ \rm -$&$ 24.7602\imo$&$20.6077\imo$&$ 4.6077\imo$&$ -$&$ -$\\
\hline
\end{tabular}
\caption{Unstable QNMs for charged GB black brane perturbation of
tensor type for fixed $\lambda$ ($\lambda=0.20$) and $k_3=500$. Note
that $a=1.4$ exceeds the maximal value of charge permitted for $9$-
and $10$-dimensional charged black brane and thus we leave the
frequency blank there.}\label{two}
\end{center}
\end{table*}
\begin{table*}[htbp]
\begin{center}
\begin{tabular}{|c|c|c|c|c|c|c|c|c|}
\hline
$\lambda$&$a=1.6$&$a=1.4$&$a=1.2$&$a=1.0$&$a=0.8$&$a=0.6$&$a=0.4$\\
\hline
$0.20$&$  131.3869\imo$&$  129.001\imo$&$122.7276 \imo$&$  110.9159\imo$&$91.1455\imo$&$59.3517\imo$&$7.0564\imo$\\
$0.15$&$ 66.3475 \imo$&$62.7713\imo$&$ 53.5830\imo$&$ 36.9280\imo$&$ 10.0682\imo$&$ \rm - $&$ \rm - $\\
$0.10$&$  16.015\imo$&$ 10.3104\imo$&$ \rm -$&$ \rm -$&$ -$&$ \rm -$&$ \rm - $\\
$0.08$&$ 0.9712\imo$&$ \rm -$&$\rm -$&$ \rm -$&$ -$&$ -$&$ \rm - $\\
\hline
\end{tabular}
\caption{Unstable QNMs for charged GB black brane perturbation of
tensor type for fixed dimensionality ($D=6$) and
$k_3=500$.}\label{three}
\end{center}
\end{table*}
In order to show explicitly the behavior of gravitational
perturbation in higher dimensions, we solve the Schr$\ddot{o}$dinger
equation (\ref{schr}) with negative valued potential numerically and
find some unstable QNMs (see tables \ref{one}, \ref{two} and
\ref{three}). Among these tables, we can find that the real part of
$\omega$ is vanishing, while the imaginary part of $\omega$ is
positive.

Table 1 demonstrates that the unstable modes of the black brane are
suppressed as $D$ increases. This confirms the result obtained in
Ref.\cite{konoplya}. From table 2 and 3, we see that lower value of
charge ($a$) and $\lambda$ stabilize the perturbation, while the
lower value of $D$ strengthens the instability. The reason for why
higher $D$ stabilize the perturbation is because that no matter how
big $D$ is,  $\lambda$ (i.e.$\lambda=(D-3)(D-4)\alpha'/l^2$) is
bounded by $1/4$ which means that for fixed $l$, $\alpha'\rightarrow
0$ as $D$ increases. Moreover, $\alpha'\rightarrow 0$ corresponds to
vanishing Gauss-Bonnet correction in (\ref{action}) and charged
black branes in that regime are stable.
\section{Conclusions and discussions}
\setcounter{equation}{0} \setcounter{footnote}{0} In summary, we
have computed the dimensionality dependence of $\eta/s$ for charged
black branes with Gauss-Bonnet correction. The ratio of the shear
viscosity to entropy density in $D$-dimensional space-time was found
to be $\eta/s=\frac{1}{4 \pi } \left(1-\frac{2\lambda}{D-3}
[(D-1)-(D-3)a]\right)$. When $D=5$, we can recover the result found
in \cite{gmsst}. It is worth noticing that for non-zero charge the
viscosity can never approaches even in 5D case.

While in \cite{fada}, it was always assumed that $\lambda \leq
\frac{1}{4}$, in this paper we have shown explicitly that  both
causality and stability constrained the value of $\lambda$ to be
bound by $1/4$ in the limit $D\rightarrow\infty$, but for fixed $D$,
these two constraints are different. It is interesting to notice
that from different physical processes (causality and stability), we
obtain the same bound. One may further check whether this is a
coincidence or not.

The instability of charged black brane with Gauss-Bonnet correction
was also analyzed in this paper. The result shows while higher value
of charge ($a$) and $\lambda$ strengthen  the perturbation, the
unstable modes of charged black brane are suppressed as $D$
increases. The suppression of the unstable modes for higher $D$ can
be explained from the fact that $\lambda$ is bounded by $1/4$ and
$\alpha~'$  approaches zero as $D$ increases.

\vspace*{10mm} \noindent
 {\large{\bf Acknowledgments}}

\vspace{1mm} We would like to thank Y. Matsuo, F. W. Shu and T.
Tsukioka for useful discussion at the early stage of this work. The
work of XHG is partly supported by Shanghai Leading Academic
Discipline Project (project number S30105). The work of SJS was
supported by KOSEF Grant R01-2007-000-10214-0. This work is also
supported by Korea Research Foundation Grant KRF-2007-314-C00052 and
SRC Program of the KOSEF through the CQUeST with grant number
R11-2005-021.

\end{document}